\documentstyle[aps,epsfig]{revtex}
\begin{document}
\tightenlines
\draft

\newcommand{\be}{\begin{eqnarray}}
\newcommand{\ee}{\end{eqnarray}}
\newcommand{\dia}{\!\!\!\!\!\!\not\,\,\,}
\newcommand{\1}{\'{\i}}

\twocolumn[\hsize\textwidth\columnwidth\hsize\csname
@twocolumnfalse\endcsname

\title{$\Lambda^0$ polarization as a probe for production of
       deconfined matter in ultra-relativistic heavy-ion collisions}  
\author{Alejandro Ayala$^1$, Eleazar Cuautle$^1$, Gerardo Herrera$^2$,
        Luis Manuel Monta\~no$^2$} 
\address{$^1$Instituto de Ciencias Nucleares,
         Universidad Nacional Aut\'onoma de M\'exico,\\
         Apartado Postal 70-543, M\'exico Distrito Federal 04510,
         M\'exico.}
\address{$^2$Centro de Investigaci\'on y de Estudios Avanzados,\\
         Apartado Postal 14-740, M\'exico Distrito Federal 07000, 
         M\'exico.}
\maketitle
\begin{abstract}
We study the polarization change of $\Lambda^0$'s produced in
ultra-relativistic heavy-ion collisions with respect to the
polarization observed in proton-proton collisions as a signal for the
formation of a Quark-Gluon Plasma (QGP). Assuming that, when the density of
participants in the collision is larger than the critical density for
QGP formation, the $\Lambda^0$ production mechanism changes from
recombination type processes to the coalescence of free valence
quarks, we find that the $\Lambda^0$ polarization depends on the
relative contribution of each process to the total number 
of $\Lambda^0$'s produced in the collision. To describe the
polarization of $\Lambda^0$'s in nuclear collisions for densities below
the critical density for the QGP formation, we use the
DeGrand-Miettinen model corrected for the effects introduced by
multiple scattering of the produced $\Lambda^0$ within the nuclear
environment. 
\end{abstract}
\vskip2pc]

\section{Introduction}

The study of the kind of matter produced during nuclear collisions at
relativistic and ultra-relativistic energies has become the subject of
an increasing experimental and theoretical effort during the last
years. The main drive for such study is the expectation that a phase
transition, from ordinary nuclear matter to a Quark Gluon Plasma
(QGP), should be observed when conditions of sufficiently high
baryonic densities and/or temperatures are achieved during the
collision. In order to identify this phase transition, a number of
experimental observables, such as $J/\Psi$ suppression, strangeness
enhancement, fluctuations in particle ratios, flow patterns, etc.,
have been proposed~\cite{Heinz}. At the same time, it has also been
recognized that no single signal can, by itself, provide clear cut
evidence for the existence of the QGP and that it is only through the
combined analysis of all possible signals that the production of such
state of matter can be firmly established. This realization has
stimulated the search for new probes to aid in the understanding of
the properties of the complex environment produced in heavy-ion
collisions at high energy.

One of the first proposed signatures to unveil the production of a QGP
in relativistic nucleus-nucleus collisions was to study the change in
the polarization properties of $\Lambda^0$ hyperons as compared to 
that observed in proton-proton
collisions~\cite{{Angert},{Panagiotou}}. At the same time, no
quantitative model has been proposed to realize such an idea in this
kind of environment.

Recall that, in high-energy proton-induced reactions, the
produced hyperons exhibit a strong polarization. Among the hyperons,
the $\Lambda^0$ plays a special role due to its rather simple spin
structure within the static quark model and to the fact that its
self-analyzing main decay mode into $p\ +\ \pi^-$, makes polarization
measurements experimentally feasible.

For proton-induced reactions, it has long been
established~\cite{Bonner} that the spin of the $\Lambda^0$ is carried
by the produced $s$-quark and that the $u$- and $d$-quarks can be
thought of as being coupled into a diquark with zero total angular
momentum and isospin. However, despite the wealth of experimental
information and theoretical insight accumulated during the last three
decades, a complete understanding of the hyperon polarization
mechanism is still missing, mainly because, as is currently believed,
this mechanism has its origin in the recombination processes whose
nature belongs to the realm of non-perturbative phenomena.

Nonetheless, a quantitative description of hyperon polarization
properties in proton- and meson-induced reactions has been attained by
a semi-classical model put forward by DeGrand and
Miettinen~\cite{DeGrand} (see also Ref.~\cite{Kubo}). In this model,
the hyperon polarization is due to a Thomas precession effect during
the quark recombination process of slow (sea) $s$-quarks and fast
(valence) $ud$-diquarks. 

In the case of relativistic nucleus-nucleus collisions, the
expectation is that, $\Lambda^0$'s coming from the zone where the
critical density for QGP formation has been achieved, are produced
through the coalescence of independent slow sea $u$-, $d$- and
$s$-quarks and are emitted via an evaporation-like
process. Consequently, these {\it plasma} created $\Lambda^0$'s should
show zero polarization~\cite{Panagiotou}.

An accurate estimate of the change of the $\Lambda^0$ polarization
observed in high-energy heavy-ion collisions as compared to
proton-proton reactions requires knowledge of the relative
contribution to the total $\Lambda^0$ yield both from the plasma zone
and from regions where the critical density for QGP formation has not
been reached. Moreover, since those $\Lambda^0$'s produced in these
latter regions experience multiple scattering with the surrounding
nuclear environment, the final polarization value should reflect the
effects introduced by these processes.

In this paper, we propose a method to extract information from
$\Lambda^0$ polarization measurements in ultra-relativistic heavy-ion
collisions as a means to determine the production of a QGP.  We study
a model where the $\Lambda^0$'s produced in the zone where the
critical density for QGP formation is reached are unpolarized while
the $\Lambda^0$'s produced in the rest of the reaction zone are
produced polarized, in the same way as they are in free
nucleon-nucleon reactions. We use the DeGrand-Miettinen Model (DMM) to
describe the polarization of these latter $\Lambda^0$'s. The effects
on the polarization arising from multiple scattering are introduced in
terms of a sequential model of collisions~\cite{{Hwa},{Csernai}}. We
also discuss possible depolarization effects introduced by spin-flip
interactions of $\Lambda^0$ within the nuclear medium.

The work is organized as follows: In Sec.~\ref{II} we briefly review
the DMM for the polarization of hyperons produced in proton-proton
reactions. In Sec.~\ref{III}, we discuss the mechanisms that can
induce a change in the polarization of hyperons produced in
high-energy nucleus-nucleus reactions as compared to proton-proton
collisions. We account for these changes in terms of a sequential
model of collisions. To illustrate our results, we introduce the
appropriate parameters which we extract from the existing (though
scarce) data on $\Lambda^0$-nucleon 
interactions. In Sec.~\ref{IV}, we present our model for the
production of $\Lambda^0$ coming from the regions in the reaction zone
with a density of participants below and above the critical density
for QGP formation. We show that the polarization of these
$\Lambda^0$'s depends on the relative contribution of each of the
above production mechanisms to the total $\Lambda^0$ yield. We also
include the modifications to the polarization introduced by multiple
scattering of $\Lambda^0$'s in those regions with a density of
participants below the critical density for QGP formation. Armed with
such expression for the $\Lambda^0$ polarization in high-energy
nucleus-nucleus reactions, we apply the formalism to study $^{208}$Pb
-- $^{208}$Pb collisions and draw quantitative conclusions about the
$\Lambda^0$ polarization behavior as a function of both its transverse
momentum and impact parameter of the reaction. We finally summarize
and conclude in Sec.~\ref{V}. In what follows, we present the
expressions in terms of quantities in the laboratory frame.

\section{$\Lambda^0$ polarization in the DeGrand-Miettinen
         model}\label{II}

The $\Lambda^0$ hyperon has spin $S=1/2$. Its preferred decay mode
$\Lambda^0 \rightarrow p +\ \pi^-$ ($64 \%$ branching ratio) is
mediated by the weak interaction, therefore, the information from the
distribution of its decay products can be used to determine the
orientation of its spin.

Since the discovery of a substantial $\Lambda^0$ transverse
polarization in high-energy, hadron-nucleon~\cite{Pepin},
nucleon-nucleon~\cite{Aahlin} and nucleon-nucleus~\cite{Bunce}
reactions, a large amount of theoretical and experimental activity has
been devoted to understanding the origin of such polarization. The
fact that $\Lambda^0$'s were produced with a significant polarization
while $\bar{\Lambda}^0$'s were unpolarized was, at first,
surprising. Several models were proposed to explain that
phenomena. One of the most successful is the DMM~\cite{DeGrand}. In
this semi-classical model, the $\Lambda^0$ polarization results from a
Thomas precession of the spin of the $s$-quark in the recombination
process. The $u$- and $d$-quarks are assumed to form a diquark in a
state with total angular momentum $J=0$ (and isospin $I=0$) and thus,
the $s$-quark is entirely responsible of the spin of the $\Lambda^0$.

The recombining quarks in the projectile must also carry a transverse
momentum given, approximately by half of the transverse momentum of
the outgoing hadron, while the other half is carried by the
$s$-quark. Since the $x$ distribution of the (sea) $s$-quarks peeks at
very low values and is very steep, a $\Lambda^0$ must get most of its
longitudinal momentum from the valence $ud$-diquark momentum.  In the
process, the $s$-quark is, on average, accelerated but, since its
transverse momentum is different from zero, the force ${\mathbf F}$
felt by the quark is not parallel to its velocity \mbox{\boldmath
$\beta$}, giving rise to Thomas precession, characterized by a frequency
\mbox{\boldmath $\omega_T$} given by \be \mbox{\boldmath $\omega_T$} =
\left(\frac{\gamma}{\gamma+1}\right) {\mathbf F} \times
\mbox{\boldmath $\beta$}\, , \label{freq} \ee where $\gamma$ is the
Lorentz gamma factor. When this frequency is used to account for the
spin-orbit term in the Hamiltonian for computing the scattering
amplitudes to produce $\Lambda^0$'s with spin oriented along and
opposite to the normal of the production plane, it gives rise to a
negative polarization, since it is more likely that the (sea)
$s$-quark is accelerated than decelerated to recombine into a
$\Lambda^0$.

Notice that in the model, the polarization asymmetry arises as a
consequence of a strong momentum ordering where the $s$-quark is (on
average) slow and the $ud$-diquark is (on average) fast. This ordering
does not happen when the recombination involves only sea $s$- $u$- or
$d$-quarks or antiquarks, as is the case of antihyperon production or,
more important to our purposes, a $\Lambda^0$ originated from a QGP.

The polarization of a $\Lambda^0$ in proton-proton collisions is given
in the DMM by \be {\mathcal P}^{\mbox {\tiny REC}}=-\frac{12}{\Delta
x_0 M^2}\left( \frac{1-3\xi (x)}{[1+3\xi (x)]^2}\right) p_{T}\, ,
\label{eq1} \ee where \be M^2=\left[\frac{m_D^2+p_{TD}^2}{1-\xi (x)}
+\frac{m_s^2+p_{Ts}^2}{\xi (x)} -m_{\Lambda^0}^2-p_{T}^2\right]
\label{eq2} \ee with $m_D$, $p_{TD}$ ($m_s$, $p_{Ts}$) the mass and
transverse momentum of the $ud$-diquark ($s$-quark), $m_{\Lambda^0}$
and $p_{T}$ the mass and transverse momentum of the $\Lambda^0$,
$\Delta x_0$ a distance scale, on the order of the proton radius,
characterizing the recombination length scale and $\xi (x)=x_s/x$ the
ratio of the longitudinal momentum fraction of the $s$-quark to the
longitudinal momentum fraction of the $\Lambda^0$ with respect to the
beam proton.

To give a quantitative description for the polarization, DeGrand and
Miettinen take a linear parametrization for $\xi (x)$ such that \be
\xi(x)=\frac{1}{3}(1-x)+0.1x\, , \label{xi} \ee which represents a
reasonable interpolation between the values for $\xi$ near $x=0$,
where the distribution for all flavors is roughly equal in shape, and
near $x=1$, where the distribution of sea and valence quarks is small,
the former being even smaller than the latter. Using this
parametrization, DeGrand and Miettinen obtain a good description of
experimental data~\cite{DeGrand}.  If, on the other hand, use is made
of a recombination model~\cite{hmms-plb} and $\xi(x)$ explicitly
computed, the results for the polarization do not show a drastic
change. We therefore use here, for the sake of simplicity, the linear
parametrization for $\xi(x)$ given by Eq.~(\ref{xi}). To complete the
overall parametrization of ${\mathcal P}^{\mbox{\tiny REC}}$, we also
take $m_D=2/3$ GeV, $m_s=1/2$ GeV and 
$p_{Ts}^2=p_{TD}^2=(1/4)p_{T}^2+\langle k_{T}^2\rangle$ with $\langle
k_{T}^2\rangle =0.25$ GeV$^2$~\cite{DeGrand}.

\section{Depolarization effects in nucleus-nucleus collisions}\label{III}

In the case of nucleus-nucleus collisions, the effects that can
possibly produce a diminishing of the $\Lambda^0$ polarization have
been enumerated in Refs.~\cite{Panagiotou}. These are:
(i) secondary $\Lambda^0$'s produced by pion-nucleon scattering, (ii)
$\Lambda^0$ production from a QGP and (iii) secondary scattering of
leading $\Lambda^0$'s with nucleons within the interaction zone.
 
Though production of $\Lambda^0$'s by pion-nucleon scattering becomes
important in collisions of large nuclei at high energy, due to both,
an increase of the pion-nucleon and the pion production cross sections
with increasing atomic number at high energies, these $\Lambda^0$'s
are mainly produced beyond the free nucleon-nucleon phase space
kinematical limit and at low (laboratory) momenta. It is thus possible
to set kinematical constraints in the reconstruction of these
$\Lambda^0$'s and therefore exclude them from the polarization
analysis. Thus, we do not further consider this effect. We postpone
the discussion on the polarization modification of $\Lambda^0$'s
originating from a QGP to Sec.~\ref{IV}. Here, we concentrate on the
effects introduced by secondary scattering of $\Lambda^0$'s with
nucleons in the zone where the density of participants is below the
threshold for a QGP formation.

Secondary scattering within nuclear matter is an important effect in
high-energy nucleus-nucleus collisions. It is responsible for the
transverse spectrum broadening of produced particles and for the
longitudinal momentum loss of nucleons, which is related to the degree
of stopping in the reaction.

Since at high energies, the trend for the existing (though scarce)
data on free $\Lambda^0$-nucleon interactions indicates that elastic
scattering dominates the total cross section~\cite{Schopper}, we
concentrate on describing the effects on the $\Lambda^0$ polarization 
by these kind of collisions.
    
Secondary elastic scattering of $\Lambda^0$'s with nucleons can
influence the final polarization measurements by producing (a) a shift
in the $\Lambda^0$ longitudinal momentum, (b) a shift in the
$\Lambda^0$ transverse momentum and (c) a flip in the original
$\Lambda^0$ spin direction. We proceed to quantitatively discuss each
one of these effects.

\subsection{Longitudinal momentum shift}   

Multiple elastic scattering can be cast in terms of a sequential
model~\cite{{Hwa},{Csernai}}, where it is assumed that the average
energy of a particle after $n+1$ collisions is given in terms of the
average energy after $n$ collisions by 
\be 
   \langle E_{\Lambda^0}
   \rangle_{n+1} = (1-I)\langle E_{\Lambda^0} \rangle_n\, ,
   \label{energy} 
\ee where $I$ is the inelasticity
coefficient~\cite{Hufner}. In the high energy limit $(p\gg
m_{\Lambda^0})$ 
\be 
   \langle E_{\Lambda^0} \rangle_n \simeq \langle
   p_{\Lambda^0} \rangle_n = p \langle x \rangle_n\, 
   \label{highlim} 
\ee
where $p$ is the initial momentum value of the nucleon that produced
(through recombination) the $\Lambda^0$ and $x$ the fraction of the
initial longitudinal $\Lambda^0$ momentum to the nucleon longitudinal
momentum. Equations~(\ref{energy}) and~(\ref{highlim}) can be combined
to find the average value of $x$ after $n$ collisions as 
\be 
   \langle x \rangle_n = (1-I)^n x\, .  
   \label{avx} 
\ee 
We now proceed to find the average value $\langle x(z,b) \rangle$,
after the produced $\Lambda^0$ has traveled a longitudinal distance
$z$, in collisions with impact parameter $b$. Recall that the
distribution probability $P_n(z,b)$ for $n$-collisions in reactions with
impact parameter $b$ is Poissonian, that is 
\be 
   P_n(z,b) =\frac{1}{n!}\bar{N}^n(z,b)e^{-\bar{N}(z,b)}\, , 
   \label{Poisson} 
\ee
with 
\be 
   \bar{N}(z,b) = \sigma_{\Lambda^0 N}^{\mbox{\tiny tot}}
   T_{A}(z,b/2)\, , 
   \label{barN} 
\ee 
being the average number of collisions after the $\Lambda^0$ has
traveled a longitudinal distance $z$, given in terms of the total
$\Lambda^0$-nucleon cross section, which we take as $\sigma_{\Lambda^0
N}^{\mbox{\tiny tot}}\simeq 25$ mb and 
\be 
   T_A(z,s)=\int_{-z/2}^{z/2}\rho_A(z',{\bf s})
   dz' 
   \label{TA} 
\ee 
being the average nucleon density per unit area in the
transverse plane with respect to the collision axis. Notice that the
argument of the function $T_A$ in Eq.~(\ref{barN}) referring to the
location of the trajectory of the $\Lambda^0$ in the transverse planes
has been taken as the geometrical average in these planes. 

Therefore, $\langle x(z,b)\rangle$ is given by 
\be 
   \langle x(z,b) \rangle &=& xe^{-\bar{N}(z,b)}
   \sum_{n=0}^\infty\frac{(1-I)^n}{n!}\bar{N}^n(z,b)\nonumber\\
   &=&xe^{-I\bar{N}(z,b)}\, .  
   \label{finlongmom} 
\ee 
It can be shown~\cite{Csernai} that $I=\lambda /2$, where 
\be 
   \lambda\simeq\frac{\sigma_{\Lambda^0 N}^{\mbox{\tiny inel}}} 
   {\sigma_{\Lambda^0 N}^{\mbox{\tiny tot}}}\, , 
   \label{lambda} 
\ee with $\sigma_{\Lambda^0
N}^{\mbox{\tiny tot}}$ and $\sigma_{\Lambda^0 N}^{\mbox{\tiny inel}}$
the total and inelastic $\Lambda^0$-nucleon cross sections,
respectively. Although these should be taken as the cross sections in
nuclear matter, such information has not been experimentally obtained
up until now. Nevertheless, from data at low and intermediate
energies, it is known that the free cross sections follow the behavior
of the corresponding nucleon-nucleon cross
sections~\cite{Hauptman}. Assuming that this is also the case within
nuclear matter, we extrapolate the existing data on free
$\Lambda^0$-nucleon interactions to high energies~\cite{Schopper} and
obtain $\lambda\simeq 0.4$ which is consistent with the corresponding
value obtained for nucleon-nucleon collisions within nuclear
matter~\cite{Csernai}.

\subsection{Transverse momentum shift}

Let $Q({\bf s_f},{\bf s_i},z)$ be the probability that the produced
$\Lambda^0$ ends up traveling in the direction ${\bf s_f}$ after
traveling a longitudinal distance $z$, having been produced moving in
direction ${\bf s_i}$. Let $q_n({\bf s_f},{\bf s_i})$ be the probability
that the $\Lambda^0$ ends up traveling in the direction ${\bf s_f}$
after $n$-collisions with nucleons, having been produced moving in
direction ${\bf s_i}$. The vectors ${\bf s_f}$
and ${\bf s_i}$ can be thought of as two dimensional unit vectors,
given the isotropy of the angular distribution in each collision.

It is easy to show~\cite{Uscinski} that the explicit expression for
$q_n({\bf s_f},{\bf s_i})$ is 
\be 
   q_n({\bf s_f},{\bf s_i})=\left(\frac{1}{n\pi\Gamma^2}\right) 
   e^{-\frac{\left({\bf s_f}-{\bf s_i}\right)^2}{n\Gamma^2}}\, , 
   \label{q} 
\ee 
where the distribution is taken as
Gaussian and for the ease of the calculations, the range of each
component of ${\bf s_f}$ and ${\bf s_f}$ is taken as
$[-\infty,\infty]$. $\Gamma$ is the average dispersion
angle in each collision. We emphasize that the expression in
Eq.~(\ref{q}) is valid for small dispersion angles. Since the
distribution probability for $n$-collisions is given by
Eq.~(\ref{Poisson}), $Q({\bf s_f},{\bf s_i},z)$ is given as 
\be 
   Q({\bf s_f},{\bf s_i},z)=
   \sum_{n=0}^\infty P_n(z,b)q_n({\bf s_f},{\bf s_i})\, .
   \label{Q} 
\ee 
The average change in the $\Lambda^0$ momentum direction
is computed from the r.m.s value of the total dispersion
angle, $\alpha$, given by 
\be 
   \alpha &=& \sqrt{\int d^2s\ {\bf s}^2\ Q({\bf s},z)}\nonumber\\ 
   &=& \Gamma\sqrt{\bar{N}(z,b)}\, ,
   \label{rms} 
\ee where ${\bf s}={\bf s_f}-{\bf s_i}$, since $Q$ depends
only on such difference. It is now a simple matter to express the
average value of the $\Lambda^0$ transverse momentum after having
traversed a longitudinal distance $z$ within the nuclear medium, as
\begin{equation}
   \langle p_T(z,b) \rangle = \left(\sqrt{p_{\Lambda^0}^2-p_T^2}
   \sin\alpha + p_T\cos\alpha\right)e^{-I\bar{N}(z,b)}\, ,
   \label{avpT}
\end{equation}
where the factor $e^{-I\bar{N}(z,b)}$ comes from considering the
average change in the magnitude of the $\Lambda^0$ momentum.

In the high-energy limit $\cos\alpha\sim 1$, and we can safely ignore
the first term in Eq.~(\ref{avpT}), writing 
\be 
   \langle p_T(z,b) \rangle \simeq
   p_Te^{-I\bar{N}(z,b)}\cos\left\{\Gamma\sqrt{\bar{N}(z,b)}\right\}\, , 
   \label{avpt2} 
\ee
where $p_T$ is the transverse momentum that the $\Lambda^0$ carried
originally and we have used Eq.~(\ref{rms}).

$\Gamma$ can be estimated using the data on
angular distributions of free $\Lambda^0$-nucleon elastic
scattering. For intermediate energies with $300\ \mbox{MeV/c} \leq
p_{\Lambda^0} \leq 1500\ \mbox{MeV/c}$, such data exists
and $\Gamma$ can be read off from the parametrization of
the angular distribution for $p_{\Lambda^0} > 800\ \mbox{MeV/c}$ in
terms of Legendre polynomials with $l=1,2,3$ in Ref.~\cite{Kadyk}. By
doing so it is easy to get that $\Gamma\simeq 1.2 $ rad., which
represents a large value. However, we expect that at
higher energies, elastic dispersion takes place with smaller values of
$\Gamma$. In fact, as the energy of the produced $\Lambda^0$ increases
from about 1 GeV to a few tenths of GeV, its deBroglie wavelength
$\bar{\lambda}$ decreases from about 0.1 fm to 0.01
fm. Since the transverse size $d$ of the scatterers (nucleons) is about 1
fm, we can estimate that~\cite{Uscinski}
$\Gamma\sim\bar{\lambda}/d=0.01$. Hereafter, we use this value of
$\Gamma$ in our calculations. 

\subsection{Spin flip}

The existing data on $\Lambda^0$ production by meson-induced reactions
on light nuclei~\cite{Ajimura} show that the effects on the
$\Lambda^0$ polarization produced by final state interactions
($\Lambda^0$-nucleon scattering) are small. Since, within a light 
nucleus, the average number of collisions experimented by a produced
$\Lambda^0$ is $\bar{N}< 1$, it is to be expected that the effects, if
any, reflect the depolarization involving the spin interactions. 

Let us first note~\cite{Conzett} that the spin-orbit interactions
cannot contribute to $\Lambda^0$ depolarization, given that this
interaction is parity-conserving. Since depolarization requires
spin-flip, the only interaction capable to produce it is the spin-spin
interaction. 

To quantify the degree of depolarization in a single
$\Lambda^0$-nucleon collision, it is customary to introduce
the polarization transfer coefficient~\cite{Bystricky} $D$, which, in the
case of forward scattering, expresses the final polarization ${\mathcal P}'$
in terms of the original polarization ${\mathcal P}$ as
\be
   {\mathcal P}'=D\ {\mathcal P}\, .
   \label{depol}
\ee
In a multiple scattering scenario, such as
the one considered here, the average depolarization due to two-body
$\Lambda^0$-nucleon interactions can be written as
\be
   \langle{\mathcal P}'(z,b)\rangle&=&{\mathcal P}\sum_{n=0}^\infty
   P_n(z,b)D^n\nonumber\\
   &=&{\mathcal P}e^{-\bar{N}(z,b)(1-D)}\, .
   \label{depexp}
\ee    
Assuming that the spin-spin interactions are isotropic, $D$ can be
expressed as 
\be
   D=\frac{\left| \bar{V} \right|^2 + \left| S_{\Lambda}\right|^2
   + \left| S_N \right|^2 - \left|\Delta\right|^2}
   {\left| \bar{V} \right|^2 + \left| S_{\Lambda}\right|^2
   + \left| S_N \right|^2 + 3\left|\Delta\right|^2}\, ,
   \label{depolcoef}
\ee
where $\bar{V}$, $S_{\Lambda}$, $S_N$ and $\Delta$ represent the
amplitudes for the spin-independent, $\Lambda^0$ spin-orbit, nucleon
spin-orbit and spin-spin interactions, respectively, appearing in the
expression for the two-body $\Lambda^0$-nucleon
potential~\cite{Millener}. 

\vspace{-0.5cm}
\begin{figure}[t!] 
\centerline{\epsfig{file=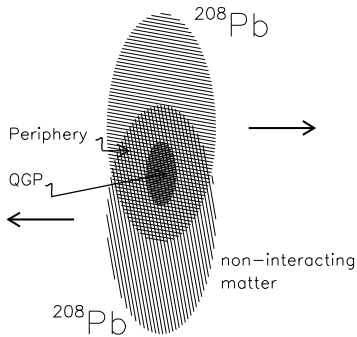,height=3.0in,width=3in}}
\vspace{-0.5cm}
\caption{The reaction $^{208}$Pb -- $^{208}$Pb and the regions
where $\Lambda^0$'s may be produced. In the QGP, $\Lambda^0$'s
originate from the QCD processes $q\bar{q} \rightarrow s \bar{s}$ 
and $gg \rightarrow s \bar{s}$. In the periphery $\Lambda^0$'s 
are produced by recombination.} 
\end{figure}

The above parameters can be reasonably constrained at low energies (a
few MeV's) from the spectral analysis of hypernuclei
levels~\cite{May}, however, no further experimental data exists at
higher energies. Nevertheless, it is clear that, when in the range of
energies of interest, the parameters appearing in
Eq.~(\ref{depolcoef}) are measured, $D$ can be considered to be a
constant smaller than unity. Therefore, for a given impact parameter,
the spin-flip interactions 
have a simple effect on the depolarization of $\Lambda^0$'s produced
in nucleus-nucleus collisions. For the purposes of this work and until
experimental information becomes available, we will omit these effects
from the analysis.     

\section{$\Lambda^0$ production and polarization in ultra-relativistic 
         heavy-ion collisions}\label{IV}

In a QGP the $s\ \bar{s}$ pair production
can be reasonably well described by the lowest order QCD processes
$q + \bar{q} \rightarrow s + \bar{s}$ and 
$g + g \rightarrow s + \bar{s}$~\cite{qcd}. To simplify the analysis,
we assume that in this environment, the $\Lambda^0$'s are the sole
products of the subsequent $s$-quark recombination. The enhancement of
strangeness production in a QGP plasma leads to an enhancement of
hyperon production. However, $\Lambda^0$ formation in this
environment should not show a strong ordering for the momenta of any
of the $u$-, $d$- or $s$-quarks, and consequently, according to the
DMM, these $\Lambda^0$'s should not be polarized~\cite{lc-plb}.

On the other hand, in the interaction region where nucleon-nucleon
collisions take place but the density is not high enough to deconfine
quarks and gluons, $\Lambda^0$'s would be produced by 
recombination of $(ud)$-diquarks, coming from the interacting nucleons
and, $s$ quarks coming from the sea.

The total cross section $\sigma_{\Lambda^0}$ for $\Lambda^0$
production is then given by these two components
\be
   \sigma_{\Lambda^0} =
   \sigma_{\Lambda^0}^{\mbox{\tiny REC}}
   +\sigma_{\Lambda^0}^{\mbox{\tiny QGP}}\, .
   \label{sigmatot}
\ee

The regions where the two distinct $\Lambda^0$ production mechanisms
take place during the collision are shown schematically in Fig.~1.

\subsection{$\Lambda^0$ production from recombination}

To describe $\Lambda^0$ production by recombination we write the 
production cross section at impact parameter $b$ in the collision of ions  
$A$ and $B$ as 
\be
   \frac{1}{\sigma_{\Lambda^0}^{NN}}
   \frac{d^2\sigma_{\Lambda^0}^{\mbox{\tiny REC}}}
   {d^2b} = T_{AB}(b)\, ,
   \label{diffsig} 
\ee
where $\sigma_{\Lambda^0}^{NN}$ is the $\Lambda^0$ production cross
section in nucleon-nucleon collisions, which we take as
$\sigma_{\Lambda^0}^{NN} = 3.2$ mb~\cite{ranft} and $T_{AB}(b)$ is
given by
\be 
   T_{AB}(z,b)=\int d^2s T_A(z,{\bf s})T_B(z,{\bf s}-{\bf b})\, ,
   \label{TAB} 
\ee 
with $T_A$ and $T_B$ given in terms of Eq.~(\ref{TA}), extending the
limits of integration over $z$ in Eq.~(\ref{TA}) to
$[-\infty,\infty]$. For $\rho_A({\bf r})$ we use the standard
Woods-Saxon density profile 
\be
   \rho_A({\bf r})= \frac{\rho_0}{1+e^{(r-R_A)/a}}\, ,
   \label{WS}
\ee
with $R_A = 1.1 A^{1/3}$ fm, $a=0.53$ fm and $\rho_0$ fixed by 
normalization 
\be
   \int\rho_A({\bf r}) d^3r = A\, ,
   \label{norm}
\ee 
giving $\rho_0=0.17$ fm$^{-3}$ for $^{208}$Pb.

We assume that each sub-collision produces final state particles
in the same way as in free nucleon reactions and thus we can estimate
the number of $\Lambda^0$'s produced by recombination from
Eq.~(\ref{TAB}). However, in order to exclude the zone where the
density of participants $n_p$ is above the critical density $n_c$ to
produce a QGP, we rewrite Eqs.~(\ref{TAB}) and~(\ref{diffsig}) as
\begin{equation}
   \frac{d^2\sigma_{\Lambda^0}^{\mbox{\tiny REC}}}{d^2b} 
   = \sigma_{\Lambda^0}^{NN}\!\!
   \int T_B({\bf b}-{\bf s})T_A({\bf s})
   \theta[n_c-n_p({\bf s},{\bf b})]d^2s,
   \label{diffsigmod}
\end{equation}
where $n_p({\bf s})$ is the density of participants at the point
${\bf s}$ in the transverse plane and $\theta$ is the step function.

The density of participants per unit transverse area during the
collision of nucleus $A$ on nucleus $B$, at an impact parameter vector
${\bf b}$, has a profile given by~\cite{blaizot}
\be
   n_p({\bf s},{\bf b})&=&
   T_A({\bf s}) [ 1-e^{-\sigma_{NN} T_B({\bf s}-{\bf b})}]\nonumber\\
   &+&
   T_B({\bf s}-{\bf b}) [ 1-e^{-\sigma_{NN} T_A({\bf s)}}]\, ,
   \label{partb}
\ee
where $\sigma_{NN}$ is the nucleon-nucleon inelastic cross
section which we take as $\sigma_{NN} = 32$ mb. The total number of
participants $N_p$ at impact parameter $b$ is 
\be
   N_p(b) = \int n_p({\bf s},{\bf b}) d^2s\, .
   \label{totalpartb}
\ee

\vspace{-0.5cm}
\begin{figure}[t!] 
\centerline{\epsfig{file=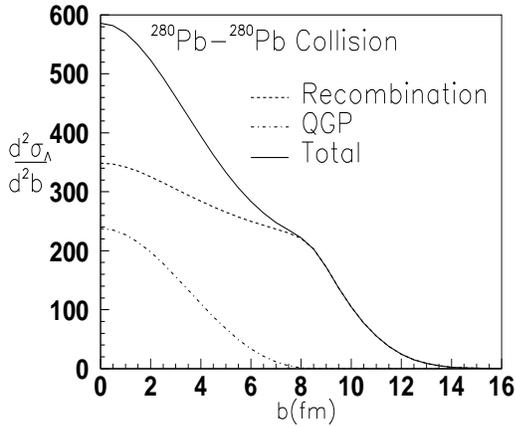,height=3.0in,width=3in}}
\vspace{-0.5cm}
\caption{$\Lambda^0$ production as a function of the impact parameter
$b$ in the QGP (dashed-dotted line) for $c=0.005$ and in the periphery
(dashed line). The solid line represents the total $\Lambda^0$ production. } 
\end{figure}

Following the reasoning in Ref.~\cite{blaizot}, we choose $n_c=3.3$
fm$^{-2}$, which results from the observation of a substantial
reduction of the $J/\psi$ yield in Pb -- Pb collisions at the
SPS. Figure~2 shows $d^2\sigma_{\Lambda^0}^{\mbox{\tiny
REC}}/d^2b$ (dashed line) as a function of $b$, computed from
Eq.~(\ref{diffsigmod}) for the case of  Pb -- Pb collisions. 

\subsection{$\Lambda^0$ production from a QGP}

The average number of strange quarks produced in a QGP scales
with the number of participants $N_p^{\mbox{\tiny QGP}}$ in the
collision roughly as~\cite{letessier} 
\be
   \frac{\langle s \rangle}{N_p^{\mbox{\tiny QGP}}} 
   = c N_p^{\mbox{\tiny QGP}}\, .
   \label{sprod}
\ee
Assuming for the sake of simplicity that, as a result of hadronization,
only $\Lambda^0$'s are obtained from these produced $s$-quarks,  
we can estimate the number of $\Lambda^0$'s originating in
the QGP. $N_p^{\mbox{\tiny QGP}}$, as a function of the impact
parameter is given, using Eq.~(\ref{totalpartb}), as
\be
   N_p^{\mbox{\tiny QGP}}(b) = \int n_p({\bf s},{\bf b})
   \theta[n_p({\bf s},{\bf b})-n_c]d^2s\, .
   \label{totalpartQGP}
\ee

\vspace{-0.5cm}
\begin{figure}[t!] 
\centerline{\epsfig{file=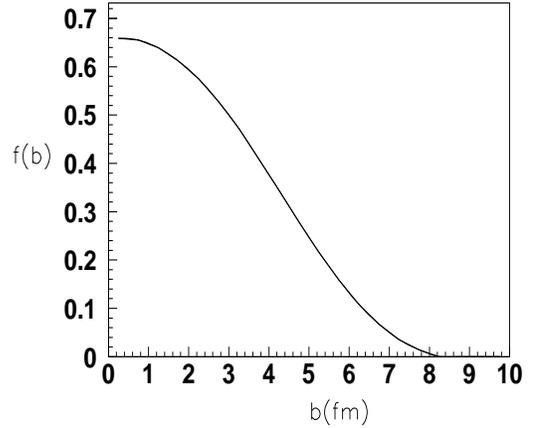,height=3.0in,width=3in}}
\vspace{-0.5cm}
\caption{The fraction $f(b)$ of $\Lambda^0$'s produced in the QGP to those 
produced by recombination as a function of impact parameter $b$.} 
\end{figure}

The proportionality constant $c$ in Eq.~(\ref{sprod}) can be read off from
Fig.~5a in Ref.~\cite{letessier}. Depending on the precise value of
$\alpha_s$ and for $^{208}$Pb -- $^{208}$Pb collisions, $c$ is found in
the range 
\be
   0.001 \leq c \leq 0.005\, .
   \label{c}
\ee
Therefore, we take Eq.~(\ref{sprod}) as representing the behavior of
the differential cross section, $d^2\sigma_{\Lambda^0}^{\mbox{\tiny
QGP}}/d^2b$, namely
\be
   \frac{d^2\sigma_{\Lambda^0}^{\mbox{\tiny QGP}}}{d^2b}=
   c \left[N_p^{\mbox{\tiny QGP}}(b)\right]^2 \, .
   \label{diffsigQGP}
\ee

Figure~2 shows $d^2\sigma_{\Lambda^0}^{\mbox{\tiny
QGP}}/d^2b$ (dashed-dotted line) as a function of $b$, computed from
Eq.~(\ref{diffsigQGP}) for the case of Pb -- Pb collisions. Shown is also the
sum $d^2\sigma_{\Lambda^0}^{\mbox{\tiny REC}}/d^2b + 
d^2\sigma_{\Lambda^0}^{\mbox{\tiny QGP}}/d^2b$ (solid line) for the
same system.

\subsection{$\Lambda^0$ polarization}

Recall that the $\Lambda^0$ polarization asymmetry ${\mathcal P}$ is
defined as the difference between the number of $\Lambda^0$'s produced with
spin pointing along and opposite to the normal of the production
plane. In terms of the differential cross sections, ${\mathcal P}$ is
given by~\cite{prd}

\begin{figure}[t!] 
\centerline{\epsfig{file=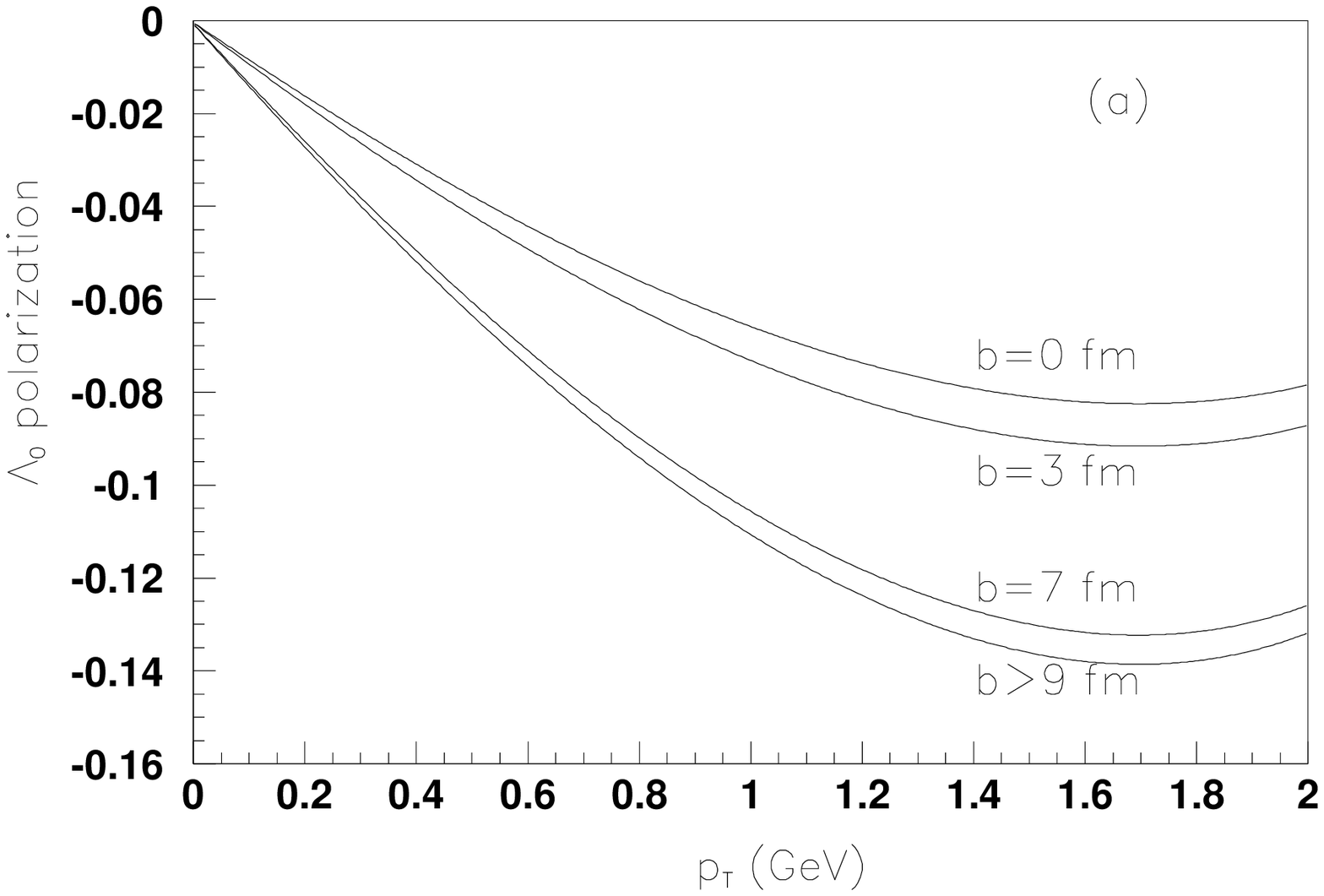,height=3.0in,width=3in}}
\vspace{-2cm}
\centerline{\epsfig{file=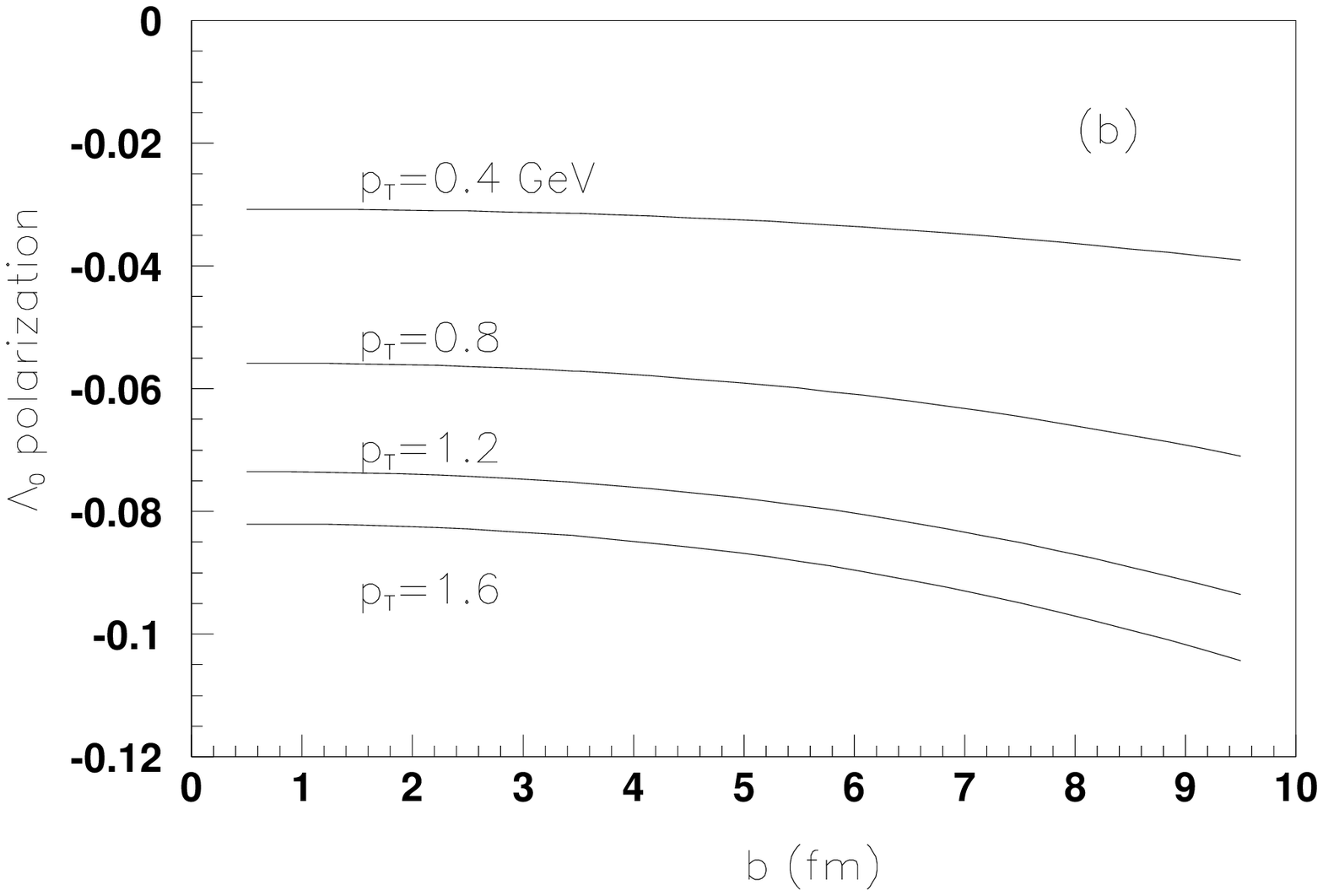,height=3.0in,width=3in}}
\vspace{-1cm}
\caption{$\Lambda^0$ polarization by recombination ($a$) as a function of $p_T$
for different values of impact parameter $b$ and (b) as a function 
of $b$ for different values of $p_T$.} 
\end{figure}

\be
   {\mathcal P}=\left[\frac{d^2\sigma_{\Lambda^0\uparrow}}{d^2b}
   -\frac{d^2\sigma_{\Lambda^0\downarrow}}{d^2b}\right]\left/
   \left[\frac{d^2\sigma_{\Lambda^0\uparrow}}{d^2b}
   +\frac{d^2\sigma_{\Lambda^0\downarrow}}{d^2b}\right.\right]\, ,
   \label{polt}
\ee
where $d^2\sigma_{\Lambda^0\uparrow}/d^2b$ and
$d^2\sigma_{\Lambda^0\downarrow}/d^2b$ are the spin up and spin down
(with respect to the normal of the production plane) differential
cross sections, respectively. In the scenario where $\Lambda^0$'s
originate from two different processes, one must take into account the
corresponding contribution to the polarization.

As we have argued, $\Lambda^0$'s coming from the QGP regions
are expected to be produced with their spins isotropically oriented and
thus these do not contribute to the net $\Lambda^0$ polarization. We
define  
\be
   f(b)=\left[\frac{d^2\sigma_{\Lambda^0}^{\mbox{\tiny QGP}}}{d^2b}\right]
   \left/
   \left[\frac{d^2\sigma_{\Lambda^0}^{\mbox{\tiny REC}}}{d^2b}\right.\right]
   \label{ratiomech}
\ee
as the ratio of the differential cross sections for $\Lambda^0$
production from the QGP and from recombination processes. Figure~3
shows $f(b)$ as a function of the impact parameter $b$, for the case
of Pb -- Pb collisions, where 
$d^2\sigma_{\Lambda^0}^{\mbox{\tiny REC}}/d^2b$ and 
$d^2\sigma_{\Lambda^0}^{\mbox{\tiny QGP}}/d^2b$ are given by
Eqs.~(\ref{diffsigmod}) and~(\ref{diffsigQGP}), respectively.

\begin{figure}[t!] 
\centerline{\epsfig{file=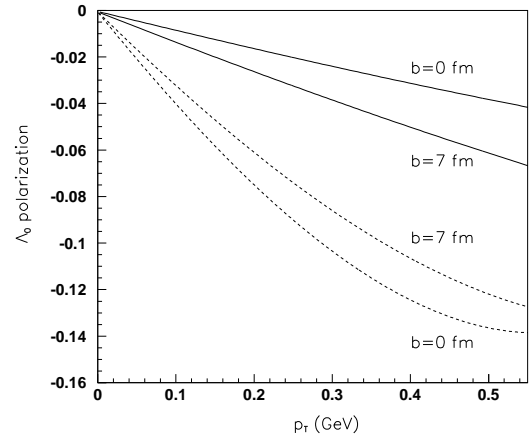,height=3.0in,width=3in}}
\vspace{-1cm}
\caption{Comparison of $\Lambda^0$ polarization values with (lower
curves) and without (upper curves) the transverse momentum shift due
to multiple scattering in the nuclear 
medium below the critical density for QGP formation.}
\end{figure}

Equation~(\ref{polt}) can be written in terms of $f(b)$ as
\be
   {\mathcal P}=\frac{{\mathcal P}^{\mbox{\tiny REC}}}
                {[1+f(b)]}\, ,
   \label{polt2}
\ee
where 
\begin{equation}
   {\mathcal P}^{\mbox{\tiny REC}}=
   \left[\frac{d^2\sigma_{\Lambda^0\uparrow}^{\mbox{\tiny REC}}}{d^2b}
   -\frac{d^2\sigma_{\Lambda^0\downarrow}^{\mbox{\tiny REC}}}
   {d^2b}\right]\left/
   \left[\frac{d^2\sigma_{\Lambda^0\uparrow}^{\mbox{\tiny REC}}}{d^2b}
   +\frac{d^2\sigma_{\Lambda^0\downarrow}^{\mbox{\tiny REC}}}
   {d^2b}\right.\right]\, ,
   \label{polrec}
\end{equation}
given in the DMM by Eq.~(\ref{eq1}), is the $\Lambda^0$ polarization
that would be observed in the absence of $\Lambda^0$'s produced by a QGP.
Figure~4 shows the polarization obtained from
Eqs.~(\ref{eq1}) and~(\ref{polt2}). Figure~4(a) shows the polarization as a 
function of $p_T$ for different values of $b$. Figure~4(b) shows the
polarization as a function of $b$ for different values of
$p_T$. 

To incorporate the effects of the shift in momentum experienced by
$\Lambda^0$'s traveling in the nuclear medium for densities below the
critical density for QGP formation, we recall that, according to the
analysis in Sec.~\ref{III}, a $\Lambda^0$ produced within a given
phase space cell labeled by the pair of values $(x,p_T)$, is scattered
into a different phase space cell, which, on average, is labeled by the
pair of values $(\langle x\rangle , \langle p_T\rangle )$. Omitting
from the analysis the spin-flip depolarization effects, the original
$\Lambda^0$ polarization computed from Eq.~(\ref{eq1}) is preserved
but corresponds (on average) to a detected $\Lambda^0$ with momenta
labels $(\langle x\rangle , \langle p_t\rangle )$. This is shown in
Fig~5 where, for the purposes of comparison, we plot the
polarization with and without considering the shift in transverse
momenta due to multiple scattering, computed with the
explcit values of $I=0.2$ and $\Gamma=0.01$ for two different
values of the impact parameter $b$.

\section{Conclusions}\label{V}

In conclusion, we propose to study the change in polarization of
$\Lambda^0$'s, with respect to proton-proton reactions, as a means to
identify the production of deconfined matter in ultra-relativistic
nucleus-nucleus collisions.

We have studied a model where $\Lambda^0$'s are produced by two
competing mechanisms, namely, recombination type of processes,
below the critical density for QGP formation, where we expect that
$\Lambda^0$'s are produced polarized, and coalescence type of
processes, above the critical density for QGP formation, where we
expect that $\Lambda^0$'s are unpolarized. The overall polarization
detected would thus depend on the relative contribution of each
process to the overall $\Lambda^0$ yield. 

To describe the polarization of $\Lambda^0$'s produced by
recombination, we use the DMM, accounting for the effects introduced
by multiple elastic scattering experienced by the produced
$\Lambda^0$'s in the nuclear environment. Multiple scattering is
responsible for two distinct effects: a momentum shift
where the polarization of the produced $\Lambda^0$'s is preserved but
their final momenta change and a depolarization due to spin flip
interactions.    

We have used the existing data on $\Lambda^0$-nucleon interactions to
obtain the inelasticity parameter $I$ and have estimated the average
dispersion angle per collision $\Gamma$. We have also given an
explicit expression for the depolarization coefficient, in terms of
the parameters describing the two-body $\Lambda^0$-nucleon
potential. These last parameters have yet to be measured at high energies.

Though our analysis is as quantitative as the existing data allow it,
there is no doubt that in order for the model to have a stronger
predictive power, more accurate data on $\Lambda^0$-nucleon
interactions at high energies are needed, which, together with an
increasing interest on finding new probes for the existence of a QGP,
can make $\Lambda^0$ polarization measurements a powerful analyzing
tool in high energy heavy-ion collisions. 

\section*{Acknowledgments}

Support for this work has been received in part by CONACyT under
an ICM grant and by DGAPA-UNAM under grant number IN108001.

\end{document}